\documentstyle[12pt]{article}
\begin{document}
\def\be{\begin{equation}}
\def\ee{\end{equation}}
\def\bea{\begin{eqnarray}}
\def\eea{\end{eqnarray}}
\newcommand\nn{\nonumber}
\def\thefootnote{\fnsymbol{footnote}}
\begin{flushright}
KANAZAWA-01-01  \\ 
NTUA-03/01 \\
February, 2001
\end{flushright}
\vspace*{0.5cm}
\begin{center}
{\Large\bf $\mu$-term due to the non-universal supersymmetry breaking
and the Higgs mass}\\
\vspace{1 cm}
{\Large  Daijiro Suematsu}
\footnote[1]{Email : suematsu@hep.s.kanazawa-u.ac.jp}
 {\Large ~and~  George Zoupanos}
\footnote[2]{Email : George.Zoupanos$@$cern.ch}
\vspace {0.7cm}\\
$^\ast${\it Institute for Theoretical Physics, Kanazawa University,\\
        Kanazawa 920-1192, JAPAN}
\vspace*{0.3cm}\\
$^\dagger${\it Physics Department, National Technical University,
Athens,\\ 
Zografou Campus, 157 80 Athens, GREECE }
    
\end{center}
\vspace{1cm}
{\Large\bf Abstract}\\  
We propose a solution to the $\mu$-problem due to the non-universal
soft supersymmetry breaking in a class of supersymmetric extensions 
of the SM with an extra U(1) and the NMSSM. 
Our scenario resembles an earlier one which has been proposed as
a solution to the $\mu$-problem in the framework of similar models but it
has rather different features from that. Theoretically crucial points in
the present scenario are the realization of the required non-universal
soft supersymmetry breaking parameters and the concern to avoid the vacuum
instability due to tadpoles.
We discuss them in some detail. In particular, on the non-universal soft 
supersymmetry breaking we discuss the superstring and the field
theory with reduction of couplings possibilities.  
We also study the weak scale radiative symmetry breaking 
in the considered models and estimate the upper mass bound 
of the lightest neutral Higgs scalar.
The latter tends to be smaller in the present scenario as compared
to the previous one.

\newpage
 \setcounter{footnote}{0}
 \def\thefootnote{\arabic{footnote}}
\noindent
{\large\bf 1.~Introduction}

 Supersymmetry is expected to solve the hierarchy problem in the standard 
 model (SM) \cite{n}. The minimal supersymmetric standard model (MSSM) 
is the simplest candidate of such models. 
As is well-known, however, there is a remnant of
the hierarchy problem called $\mu$-problem \cite{mu}. 
 $\mu$ is a supersymmetric mass parameter in the MSSM and it needs 
 to be close to the weak scale for the realization of a correct vacuum. 
In principle there is no reason why it should be as the weak scale and
 it is natural for such a supersymmetric mass parameter to be of the
 order of the unification scale or the Planck scale.
 
 Natural solutions for this problem may be obtained by connecting it with the
 soft supersymmetry breaking scale. Several scenarios have been proposed
 in this direction by now [3-10]. 
 One of the representative scenarios is to replace a $\mu$-term,
 $\mu H_1H_2$ with a Yukawa type coupling $\lambda SH_1H_2$ by
 introducing an SM singlet chiral superfield $S$ .
This kind of proposal requires also the introduction of some symmetry to 
 prohibit a bare $\mu$-term. 
The next to the minimal supersymmetric standard model (NMSSM) 
with a suitable discrete symmetry is an example of such scenarios. 
In fact, the radiative correction induces a weak scale 
 vacuum expectation value of the scalar component $\tilde S$ of $S$ and
the $\mu$-scale is realized as 
$\lambda\langle \tilde S\rangle$ \cite{singlet,color}.
 However, in this model the spontaneous breaking of the discrete
symmetry creates the cosmologically dangerous domain wall problem.
In addition to the bare $\mu$-term problem, 
the model is also annoyed by the vacuum instability problem 
if the discrete symmetry is not imposed \cite{nonren}. 
 
 We can consider the above proposed symmetry as a gauge symmetry.
 It has been known that a certain kind of supersymmetric extensions 
 of the SM involving an extra U(1) 
can solve the $\mu$-problem by replacing the $\mu$-term to a 
Yukawa coupling
among a SM singlet field $S$ and the usual Higgs doublet
 fields such as $\lambda SH_1H_2$ \cite{sy,extra}. 
 The bare $\mu$-term is forbidden by this extra U(1)-symmetry 
 if $S$ carries non-trivial U(1)-charge.
 Thus the superpotential of this model has the form 
 \begin{equation}
 W=\lambda SH_1H_2 + {(\rm MSSM~Yukawa~couplings)} + \cdots.
\label{eqa}
 \end{equation}
Moreover it has been checked that radiative breaking of this 
U(1)-symmetry could occur
 by making the soft supersymmetry breaking scalar squared mass 
of $\tilde S$ negative ($m_S^2 <0$), at least if there were 
vector like extra colored chiral 
 superfields $(g, \bar g)$ coupled with $S$ in the form  
$kSg\bar g$ \cite{sy,cde,ds}.
As in the case of the NMSSM, the coupling is realized as 
$\lambda\langle\tilde S\rangle$.
 A corresponding soft supersymmetry breaking parameter $B$ is also 
 induced as a usual trilinear soft supersymmetry
 breaking parameter $A_\lambda$ corresponding to the 
$\lambda SH_1H_2$ term in $W$. 
 
 In this paper we propose a new scenario for solving this problem based on the
 non-universality of the soft supersymmetry breaking parameters.
 This scenario is very similar to the one proposed previously 
in the class of models with an extra U(1) but potentially
it can have different features from it. 
In the general framework in which supergravity describes the low energy
effective theory of superstring theories the supersymmetry breaking
parameters can appear in a non-universal form \cite{il,kl,bim}.
Also a class of
renormalizable N=1 supersymmetric gauge theories, 
in which renormalization
group invariant (RGI) relations among couplings have been assumed, have
naturally similar features. 
 If soft supersymmetry breaking scalar squared masses are non-universal 
and some of them are negative, we can consider the possibility
that the breaking of the symmetry, which protects the appearance of a bare
$\mu$-term, has different features, for example, on the Higgs mass
bound. It might provide the origin of the $\mu$-scale. 

 Let us consider the model with the same superpotential as in eq.~(\ref{eqa}).
 The scalar potential of $\tilde S$ is composed only of the extra 
U(1) D-term and 
the soft breaking scalar mass term such as 
 \begin{eqnarray}
 V={g_X^2\over 2}Q_S^2 |\tilde S|^4 + \tilde m_S^2|\tilde S|^2,
\label{eqb}
 \end{eqnarray}
where $g_X$ is the extra U(1) gauge coupling and $Q_S$ is the
corresponding charge of $S$.
If $\tilde m_S^2<0$ is radiatively realized at some scale satisfying 
 the positiveness of other masses, the scalar component of $S$
obtains the vacuum 
 expectation value $|\langle\tilde S\rangle|^2=-\tilde m_S^2/(g_XQ_X)^2$. 
In this case the vacuum expectation value 
$\langle\tilde S\rangle$ is always expected to 
 be around the supersymmetry breaking 
 scale. Even if the negative $\tilde m_S^2$ is introduced 
as a result of non-universal 
 supersymmetry breaking at the unification scale $M_X$, 
this estimation will be kept. 
However in the latter case the $\mu$-term may be considered to be 
induced at the unification 
scale as a direct result of the non-universality of the soft 
supersymmetry breaking.
In order to find out the features arising in such a 
case at the weak scale we need a study using the renormalization group 
equations.
The components of the extra U(1)-vector superfields
do not decouple until the weak scale. So the phenomenology at the weak
scale may be very similar to the usual low energy breaking scenario
except that there is, in principle, no need to introduce 
large coupling of 
extra colored fields $g$ and $\bar g$ to $S$ and therefore the result
of the running of the
renormalization group equations (RGEs) is different from the 
usual case \cite{sy,extra}.

Crucial points for the success of the present scenario will be 
a consistent realization of the non-universal soft scalar masses 
required and a solution of the problem of the 
vacuum instability due to the non-renormalizable interaction 
terms of $S$ in the frameworks that they appear \cite{nonren}.
It is also interesting to examine what kind of phenomenological 
differences appear between
the $\tilde m_S^2>0$ case and the $\tilde m_S^2<0$ case.
The same scenario is applicable to the NMSSM.

The organization of this paper is as follows. 
In section 2 we will briefly review
the general structure of the soft supersymmetry breaking parameters
in the effective supergravity of superstring.
In section 3 we will present some examples which are expected to realize 
the above scenario and discuss some important points concerning the
relevant models.
In section 4 we present the field theoretical framework that
accommodates the non-universal soft supersymmetry breaking scalar mass
squared terms.
In section 5 the mass bound of the lightest neutral Higgs 
for the model with an extra U(1) and the NMSSM is discussed
as a representative example of the features that discriminate 
this class of models from others.
Section 6 is devoted to the summary. \\

\noindent 
{\large\bf 2.~Non-universal supersymmetry breaking in the effective
supergravity of superstring}

 Non-universality of the soft scalar mass is an essential point in our 
 scenario. It will be useful to start with a brief review of the gravity
 mediated supersymmetry breaking \cite{il,kl,bim}.  
 Low energy effective supergravity theory of superstring is characterized 
 in terms of the K\"ahler potential $K$, the superpotential $W$ and 
 the gauge kinetic function $f_a$.
Each of these is a function of ordinary massless chiral matter 
 superfields $\Psi^I$ and gauge singlet fields $\Phi^i$ called 
moduli \footnote{
 We are using the terminology ``moduli'' in the generalized meaning.
 A dilaton ${\cal S}(\equiv \Phi^0)$ is included in $\Phi^i$ other than the
 usual moduli $M^i(\equiv\Phi^i~(1\le i\le N))$.}, whose potential is 
perturbatively flat as far as the supersymmetry 
 is unbroken.
 We assume here that the non-perturbative phenomena such as 
 gaugino condensation occur in a hidden sector.
 After integrating out the fields relevant to these phenomena,
 the K\"ahler potential and the superpotential are expanded in the low 
 energy observable matter fields $\Psi^I$ as\footnote{Here we do not
 assume the existence of the effective $\mu$-term in W \cite{cm} and also 
 the Giudice-Masiero term in $K$ which can become the origin of 
$\mu$-term \cite{gm}.},
 \begin{eqnarray}
 &&K=\kappa^{-2}\hat K(\Phi,\bar \Phi)+
 Z(\Phi,\bar \Phi)_{I{\bar J}}\Psi^I\bar \Psi^{\bar J}+\cdots, 
\label{eqc} \\
 &&W=\hat W(\Phi)+{1 \over 3}\tilde h(\Phi)_{IJK}\Psi^I\Psi^J\Psi^K +\cdots,
\label{eqd}
 \end{eqnarray}
 where $\kappa^2=8\pi/M_{\rm pl}^2$.
 The ellipses stand for the terms of higher orders in $\Psi^I$.
 In eq.~(4), $\hat W(\Phi)$ is considered to be
 induced by the non-perturbative effects in the hidden sector.
 Using these functions the scalar potential $V$ can be written as \cite{csgp},
 \begin{equation}
 V=\kappa^{-2}e^G\left[G_\alpha(G^{-1})^{\alpha \bar \beta}G_{\bar \beta}
 -3\kappa^{-2}\right]+({\rm D-term}),
\label{eqe}
 \end{equation}
 where $G=K+\kappa^{-2}\log \kappa^6 |W|^2$ and the indices $\alpha$ and
 $\beta$ denote $\Psi^I$ as well as $\Phi^i$.
 The gravitino mass $m_{3/2}$ which characterizes the scale of supersymmetry 
 breaking is expressed as
 $m_{3/2}=\kappa^2e^{\hat K/2}|\hat W|$.
 In order to get the soft supersymmetry breaking terms in the low 
 energy effective theory from eq.~(\ref{eqe}), we take the flat limit 
 $M_{\rm pl}\rightarrow \infty$ preserving $m_{3/2}$ fixed.
 Through this procedure we obtain the effective renormalizable 
 superpotential $W^{\rm eff}$
 and the soft supersymmetry breaking terms ${\cal L}_{\rm soft}$.
 However, in the real world $M_{\rm pl}\not= \infty$ 
and then we have a tower of
 non-renormalizable terms suppressed by powers of $M_{\rm pl}$.
 These non-renormalizable terms could induce a vacuum instability as
 discussed in Ref. \cite{nonren}. We will come back to this point later.
 
 In the effective superpotential $W^{\rm eff}$ Yukawa couplings are 
 rescaled as $h_{IJK}= e^{\hat K/2}\tilde h_{IJK}$.
 The soft breaking terms ${\cal L}_{\rm soft}$ corresponding to $W_{\rm eff}$
 is defined by
 \begin{equation}
 {\cal L}_{\rm soft}=-\tilde m^2_{I\bar J}\psi^I\bar\psi^{\bar J}
 -\left({1\over 3}A_{IJK}\psi^I\psi^J\psi^K + {\rm h.c.}\right),
\label{eqf}
 \end{equation}
 where $\psi^I$ represents the scalar component of $\Psi^I$.
 Bilinear terms of $\Psi_I$ are assumed to be forbidden by some symmetry.
 Each soft breaking parameter is expressed by using $K$ and $W$ as 
 follows \cite{kl}\footnote{These soft breaking parameters are not 
 canonically normalized because a kinetic term of $\psi^I$ is expressed as 
 $Z_{I\bar J}\partial^\mu\psi^I\partial_\mu\bar\psi^{\bar J}$. },
 \begin{eqnarray}
 &&\tilde m_{I{\bar J}}^2=m_{3/2}^2Z_{I{\bar J}}
 -F^i{\bar F}^{\bar j}\left[\partial_i\partial_{\bar j}Z_{I{\bar J}}
 -\left(\partial_{\bar j}Z_{N{\bar J}}\right)Z^{N{\bar L}}
 \left(\partial_iZ_{I{\bar L}}\right)\right]
 +\kappa^2V_0Z_{I{\bar J}}, \label{eqg} \\
 &&A_{IJK}=F^i\left[\left(\partial_i +{1 \over 2}{\hat K}_i\right)h _{IJK}
 -Z^{\bar ML}\partial_iZ_{\bar M(I} h_{JK)L}\right], \label{eqh} 
 \end{eqnarray}
 where $F^i$ is an F-term of $\Phi^i$ and $\partial_i$ denotes
 $\partial/\partial \Phi^i$.
 $V_0$ is the cosmological constant expressed as
 $V_0=\kappa^{-2}(F^i{\bar F}^{\bar j} \partial_i \partial_{\bar j}
 \hat K-3m_{3/2}^2)$.
 From these expressions we find that the soft breaking parameters are 
 generally non-universal 
 and their structure is dependent on the form of K\" ahler potential, 
 especially, the functional form of $Z_{I\bar J}$.
The canonically normalized gaugino mass is known to be written by 
using the gauge kinetic function $f_a(\Phi)$ as \cite{csgp}
\begin{equation}
M_a={1\over 2}({\rm Re} f_a)^{-1}F^i\partial_i f_a.
\label{eqi}
\end{equation}
 
 Now we apply the above formalism to our model.
 The chiral superfields $\Psi^{I}$ represent quarks and leptons 
 $Q^\alpha, \bar U^\alpha, \bar D^\alpha,
 L^\alpha$ and $\bar E^\alpha$ {\it etc.} where $\alpha$ is a 
generation index. 
Then, the effective superpotential $W^{\rm eff}$ and
 soft supersymmetry breaking terms ${\cal L}_{\rm soft}$ in our model
 can be written as
 \begin{equation}
 \label{eqj}
 W^{\rm eff}= h^U_{\alpha\beta H_2}\bar U^\alpha H_2Q^\beta 
 + h^D_{\alpha\beta H_1}\bar D^\alpha H_1Q^\beta 
 + h^E_{\alpha\beta H_1}\bar E^\alpha H_1L^\beta +\lambda SH_1H_2
  +kSg\bar g,
 \end{equation}
 \begin{eqnarray}
 \label{eqk}
 {\cal L}_{\rm soft}&=&-\sum_{I,J} z^{I\dag} \tilde m_{\bar IJ}^2z^J  
 -\left(A^U_{\alpha\beta H_2}\bar U^\alpha H_2Q^\beta 
 +A^D_{\alpha\beta H_1}\bar D^\alpha H_1D^\beta 
 +A^E_{\alpha\beta H_1}\bar E^\alpha H_1L^\beta 
 \right.\nonumber \\
 && \qquad\qquad\qquad
 +\left. A_\lambda SH_1H_2+ A_kSg\bar g +{\rm h.c.}\right),
 \end{eqnarray}
 where we do not consider the gaugino masses which are irrelevant in our
 present discussion.
 The bare $\mu$-term in $W^{\rm eff}$ is forbidden by the extra U(1)-symmetry.
 The first term of eq.~(\ref{eqk}) represents the mass terms of all scalar 
 components of chiral superfields $(z^I=Q^\alpha, U^\alpha, 
 D^\alpha, L^\alpha, E^\alpha, H_1, H_2, S, g, \bar g)$ in our model. 
 The first place to check the validity of our scenario is 
 to study whether we can construct the model in which the soft scalar
 masses are consistently realized in the required way such as
 \begin{equation} 
 \tilde m_S^2 < 0, \qquad \tilde m_{H_1}^2 >0, \qquad 
\tilde m_{H_2}^2 >0,\qquad \tilde m_{\hat I}^2 >0,
\label{eql}
 \end{equation}
 where $\hat I =Q,\bar U, \bar D, L, \bar E, g, \bar g $.
 As seen from the general expressions of soft breaking parameters 
 in eqs.~(\ref{eqg}) and (\ref{eqh}), their structure 
is determined by the moduli dependence 
 of $Z_{I\bar J}$ and $W$\footnote{Only known exception is the dilaton 
 dominated supersymmetry breaking. Our scenario cannot be applicable to
 this case.}.
 
 In order to apply the above general results to construct a model satisfying 
 eq.~(\ref{eql}) and proceed further investigation 
of the validity of our scenario,
 it is necessary to make the model more definite by introducing some 
 assumptions \cite{sum}. 
 We consider a concrete model within the framework which
 satisfies the following conditions.
 At first we assume the simplest target space duality $SL(2,{\bf Z})$ 
 invariance \cite{il}
 \begin{equation}
 M^i \rightarrow {a_iM^i-ib_i \over ic_iM^i+d_i} \qquad 
 (a_id_i-b_ic_i=1 , ~a_i,b_i,c_i,d_i \in {\bf Z}),
\label{eqm}
 \end{equation}
 for each usual modulus $M^i$. 
 Under the target space duality transformations (13) the chiral superfields 
 $\Psi^I$ are assumed to transform as
 \begin{equation}
 \Psi^I \rightarrow \left(ic_iM^i +d_i\right)^{n^i_I}\Psi^I,
\label{eqn}
 \end{equation}
 where $n^i_I$ is called the modular weight and takes a suitable rational 
 value.
 This requirement also results in the invariance under the 
 following K\"ahler transformation,
 \begin{equation}
 K \rightarrow K +f(M^i)+\bar f(\bar M^{\bar i}),\qquad
 W \rightarrow e^{-f(M^i)}W.
\label{eqo}
 \end{equation}
 Additionally K\"ahler metric and then the kinetic terms of chiral superfields
 are assumed to be flavor diagonal 
 $Z_{I\bar J}=Z_I\delta_{IJ}.$
 This is satisfied almost in all known superstring models
 as suggested in \cite{bim}.
 
 In order to parametrize the direction of supersymmetry breaking in the moduli
 space,
 we introduce the parameters $\Theta_i$ which correspond to the generalized
 Goldstino angles in the moduli space \cite{bim,ksyy}.
 They are defined as
 \begin{equation}
 F^i\sqrt{\hat K_{i\bar j}}=\sqrt 3Cm_{3/2}\Theta_i,
 \qquad \sum_{i=0}^N\Theta^2_i =1,
\label{eqp}
 \end{equation}
 where we take the $\kappa=1$ unit. $N$ is the number of usual moduli 
 $M^i$ in the models.
 A constant $C$ satisfies $V_0=3\kappa^{-2}(|C|^2-1)m_{3/2}^2$.
 In the followings we assume $C=1$ and then $V_0=0$.
 The introduction of these parameters makes it possible to discuss the soft 
 breaking parameters without refering to the origin of supersymmetry breaking.
 
 In the presently known perturbative superstring models the $\hat K$ can be 
 generally written as
 \begin{equation}
 \hat K=-\sum_{i=0}^N\ln\left(\Phi^i +\bar \Phi^{\bar i}\right).
\label{eqr}
 \end{equation}
 If we apply the above assumptions to $Z_{I\bar J}$ in the K\"ahler
 potential $K$, we can constrain the functional form of $Z_I$ as
 \begin{equation}
 Z_I=\prod_{i=1}^N (M^i +\bar M^{\bar i})^{n^i_I}.
\label{eqs}
 \end{equation}
 Using these facts in eqs.~(\ref{eqg}) and (\ref{eqh}) and 
normalizing them canonically,
 we can write down the soft breaking 
 parameters as \footnote{It should be noted that these are the tree level
 results. However, the introduction of string one-loop effects \cite{one}
 will not change the qualitative features discussed here. }
 \begin{eqnarray}
 &&\tilde m_{I}^2=m_{3/2}^2\left( 1 
 +3\sum_{i=1}^N\Theta_i^2n_{I}^i \right),  \label{eqt}\\
 &&{A_{\hat I\hat JH_\alpha}\over h_{\hat I\hat JH_\alpha}}
 =-\sqrt 3m_{3/2}\sum_{i=0}^N\Theta_i\left[n_{\hat I}^i+n_{\hat J}^i
 +n_{H_{\alpha}}^i+1 - (M_i+M_i^*){\partial_i \tilde 
h_{\hat I\hat JH_\alpha}\over \tilde h_{\hat I\hat JH_\alpha} }\right], 
\label{equ}\\ 
 && {A_\lambda\over\lambda}=-\sqrt 3m_{3/2}
 \sum_{i=0}^N\Theta_i\left[n_S^i+n_{H_1}^i+n_{H_2}^i+1 
 - (M_i+M_i^*){\partial_i \tilde\lambda\over \tilde\lambda }\right],
\label{eqv} 
 \end{eqnarray}
 where $H_\alpha$ stands for $H_1$ and $H_2$.
An index $I$ represents all chiral superfields and $\hat I$ 
 and $\hat J$ represent quarks and leptons $Q_\alpha, \bar U_\alpha,
 \bar D_\alpha, L_\alpha$ and $\bar E_\alpha (\alpha =1 \sim 3)$.
Here we do not display the expression of $A_k$ which is similar to the one of 
other $A$-parameters.
 In these formulae $n_I^0=0$ since we do not consider 
 the transformations such as eqs.~(\ref{eqm}) and (\ref{eqn}) for a dilaton.
There is the gaugino mass as a remaining soft supersymmetry breaking 
parameter. Recalling that the gauge kinetic function is known that can  
be written as $f_a = k_a{\cal S}$ and using eq.~(\ref{eqi}), we obtain   
\begin{equation}
M_a=\sqrt 3 \Theta_0m_{3/2},
\label{eqw}
\end{equation}
where we neglect the threshold correction to $f_a$.

 Taking account of the functional form of $\hat K$ in eq.~(\ref{eqr}), 
 our assumption requires $f(M^i)=-\ln(ic_iM^i+d_i)$ in eq.~(\ref{eqo}).
 As a result, eq.~(\ref{eqo}) shows that the transformation property of
 superpotential $W$ is:
 $W \rightarrow \prod_{i=1}^N\left( ic_iM^i +d_i\right)^{-1}W.$ 
 Thus its coefficient function $\tilde h_{IJK}$
 transforms as the modular forms under the duality transformation of 
 the moduli fields $M^i$,
 \begin{equation}
 \tilde h_{IJK} \rightarrow \prod_{i=1}^N\left( 
 ic_iM^i +d_i\right)^{-(n_I^i+n_J^i+n_K^i+1)} \tilde h_{IJK}. 
\label{eqx}
 \end{equation}
 This shows that $\tilde h_{IJK}$ is a modular function of $M_i$ with 
 modular weight the $-(n_I^i+n_J^i+n_K^i+1)$.
 If coefficient functions $\tilde h_{IJK}$ of
 the superpotential are independent of the moduli fields whose F-terms 
 contribute to the supersymmetry breaking:
 $\partial_i\tilde h_{IJK}= 0$ for $\langle F^i\rangle \not=0$, 
 then the superpotential $W$, except for a $\hat W$ part,
 can depend only on moduli which do not contribute the supersymmetry 
 breaking.  Then the last terms in eqs.~(\ref{equ}) and (\ref{eqv}) 
have no contribution\footnote{In this special situation 
such Yukawa couplings $\tilde h_{IJK}$ 
 can be dynamical variables at the low energy region as discussed 
 in Ref.\cite{bd}. It may be a phenomenologically interesting case.}.
In such a case the relation
 \begin{equation}
 n_I^i+n_J^i+n_K^i+1=0, 
\label{eqy}
 \end{equation}
 should be fulfilled for each $i(\not=0)$. 
When this condition is satisfied, we can easily check that 
one of the sum rules \cite{sum,bims} is satisfied as
\begin{equation}
\tilde m_I^2 +\tilde m_J^2 +\tilde m_K^2 = 
\left({A_{IJK}\over h_{IJK}}\right)^2=M_a^2.
\label{eqz}
\end{equation}
 However, even if $\tilde h_{IJK}$ is a nontrivial modular function of
 $M_i$ whose auxiliary component has non-zero VEV $\langle F^i\rangle \not=0$, 
such a situation as $\partial_i\tilde h_{IJK} \ll \tilde h_{IJK}$ can happen. 
In such a case the last term in eqs.~(\ref{equ}) and (\ref{eqv}) can be
 neglected again, although the sum rule (\ref{eqz}) may not be
satisfied unless we consider the suitable supersymmetry breaking 
direction $\Theta_i$.
Although we assume that the last term in eqs.~(\ref{equ})
 and (\ref{eqv}) is negligible in the following discussion, 
we will take account of both possibilities on this sum rule.\\

\noindent
{\large\bf 3.~Negative scalar mass squared}

In the previous section we saw that the structure of 
soft supersymmetry breaking parameters are determined by the 
supersymmetry breaking direction $\Theta_i$ and 
modular weights of the matter fields. 
If we use these formulas, we can write down the 
required condition (\ref{eql}) for them to realize our scenario consistently. 
It is rewritten as
 \begin{equation}
 \sum_{i=1}^N\Theta^2_i n_S^i< -{1 \over 3}, \qquad 
 \sum_{i=1}^N\Theta^2_i n_{H_\alpha}^i > -{1 \over 3}, \qquad 
 \sum_{i=1}^N\Theta^2_i n_{\hat I}^i > -{1 \over 3},
\label{eqaa}
 \end{equation}
 where $\hat I$ stands for the fields including the MSSM contents 
other than $H_\alpha$ and $S$.
It is instructive to give some concrete examples 
which satisfy this condition in the framework of 
perturbative superstring.
As representative examples, we consider two models which have 
an overall modulus and multi moduli, respectively.

The first example is a model with an overall modulus $(N=1)$.
In this case the soft scalar mass formula reduces to 
 $\tilde m_{I}^2=m_{3/2}^2\left( 1 +\Theta_1^2n_{I} \right)$.
If all the matter fields have $n_I=-1$, we have universal soft
supersymmetry breaking parameters such as
\begin{equation}
\tilde m_I^2=m_{3/2}^2\Theta_0^2, \qquad  
{A_{IJK}\over h_{IJK}}
=-\sqrt 3m_{3/2}\Theta_0,\qquad
M_a=\sqrt 3m_{3/2}\Theta_0.
\label{eqab}
\end{equation}
This structure is the same as in 
the celebrated dilaton dominated case. Now we consider the necessary change
to realize our scenario.
The condition (\ref{eqaa}) becomes $\Theta^2_1n_S<-1$ for $S$ and
 $\Theta^2_1n_{\hat I}>-1$ for other fields including the MSSM contents.
In order to satisfy them, we must modify the modular weight of $S$ at least.
As the simplest modification we change the modular weight of $S$ 
into $n_S=-2$. The universality among the soft scalar masses of 
the ordinary MSSM contents is kept by this change.
The soft scalar masses and $A$-parameters are shifted into
\begin{eqnarray} 
&&\tilde m_S^2=m_{3/2}^2(\Theta_0^2-\Theta_1^2), \qquad 
\tilde m_{\hat I}^2=m_{3/2}^2\Theta_0^2,  \nonumber \\
&&{A_\lambda\over \lambda}=-\sqrt 3m_{3/2}\left(\Theta_0
-{\Theta_1\over 3}\right), \qquad
{A_{\hat I\hat J\hat K}\over h_{\hat I\hat J\hat K}}=
-\sqrt 3m_{3/2}\Theta_0,   
\label{eqac}
\end{eqnarray}
where the suffix $\hat I$ includes the ordinary Higgs doublet fields.
The $A$-parameter corresponding to $kSg\bar g$ term has the same
expression as $A_\lambda$.
In this situation our scenario will be realized as far as 
$1/2<\Theta_1^2< 1$. 
If we tend $\Theta_1^2$ to one, $\tilde m_S^2$ becomes much more negative.
In this example the condition (\ref{eqy}) is satisfied in the ordinary MSSM 
Yukawa couplings.
On the other hand, since the $\lambda$-term does not satisfy it,
there appears a $\Theta_1$ dependence as can be easily seen 
from eq.~(\ref{eqv}).
In general, for the ordinary Yukawa couplings in the superpotential
it is well-known that the following condition on the soft
supersymmetry breaking parameters is required to prohibit the 
undesired symmetry breaking \cite{color}\footnote{
Of course this condition should be satisfied at the weak scale.}:
\begin{equation}
\tilde m_I^2 +\tilde m_J^2 +\tilde m_K^2 >{1\over 3}\left({ A_{IJK} 
 \over h_{IJK}}\right)^2.
\label{eqad}
\end{equation} 
We should check the validity of this condition 
for the ordinary MSSM Yukawa couplings to
quarantee the consistency of our scenario. However,
it is trivially satisfied at least at the unification scale 
since the sum rule (\ref{eqz}) holds in this example.
 
As a next example, we adopt the $Z_2\times Z_2$ orbifold with 
three independent moduli $(T_1,T_2,T_3)$ \cite{il}.
In this orbifold we can consider a model composed of the fields in the 
untwisted sector with modular weights $(-1, 0, 0)$, $(0, -1, 0)$, $(0, 0, -1)$
and the fields in the twisted sector with modular weight $(-1/2, -1/2, 0)$.
Here we assume that the modular weights of $H_1$, $H_2$ and $S$ are  
$(-1, 0, 0)$, $(0, -1, 0)$ and $(0, 0, -1)$ respectively and other
matter contents
in the MSSM and $(g, \bar g)$ have $(-1/2, -1/2, 0)$. 
This assignment guarantees the
universality of the soft scalar masses among the MSSM contents except
for the doublet Higgs fields.
Under this assumption we can write down the soft supersymmetry breaking
parameters $\tilde m_I^2$ and $A_{IJK}$ as follows,
\begin{eqnarray}
&&\tilde m_{\hat I}^2=m_{3/2}^2\left(1-{3\over 2}
(\Theta_1^2+\Theta_2^2)\right), \quad
\tilde m_{H_{\alpha}}^2=m_{3/2}^2\left(1-3\Theta_{\alpha}^2\right), \quad
\tilde m_{S}^2=m_{3/2}^2\left(1-3\Theta_3^2\right),  \nonumber \\
&&{A_{\hat I\hat JH_{\alpha}}\over  
h_{\hat I\hat JH_{\alpha}}}=-\sqrt 3m_{3/2}
\left(\Theta_0-\Theta_{\alpha}+\Theta_3\right), \qquad
{A_\lambda\over \lambda}={A_k\over k}=-\sqrt 3m_{3/2}\Theta_0,
\label{eqae}
\end{eqnarray}
where the index $\alpha$ takes 1 and 2. 
$A_\lambda$ corresponds to the $B$-parameter in the MSSM.
At this stage the structure of soft
supersymmetry breaking parameters is completely 
determined by the supersymmetry breaking direction $\Theta_i$.
Here we might note that eq.~(\ref{eqy}) is satisfied only in the $\lambda$-term
in this modular weight assignment.
The ordinary MSSM Yukawa couplings do not satisfy eq.~(\ref{eqy}) and then
it is not trivial whether the condition (\ref{eqad}) can be satisfied or not.
The breaking direction $\Theta_i$ should be constrained 
by requiring to satisfy it. 
It is useful to note that the gaugino mass and then 
$\tilde m_{H_1}^2+\tilde m_{H_2}^2+\tilde m_{S}^2$ becomes 
smaller by making $\Theta_0$ smaller. 
 
In this example the supersymmetry breaking pattern similar to 
the dilaton dominance is realized by $\Theta=(1/\sqrt 3, -\sqrt 2/3, 
-\sqrt 2/3, -\sqrt 2/3)$ 
and each parameter is represented by
\begin{eqnarray}
&&\tilde m_{\hat I}^2=\tilde m_{H_\alpha}^2=\tilde m_S^2={1\over 3}m_{3/2}^2, 
\qquad M_a=m_{3/2},\nonumber \\
&&{A_{\hat I\hat JH_\alpha}\over h_{\hat I\hat JH_\alpha}}=
{A_\lambda\over \lambda}={A_k\over k}=-m_{3/2}.
\label{eqaff}
\end{eqnarray}
This satisfies the sum rule (\ref{eqz}). If we change $\Theta_i$ from this
case, we can introduce the non-universality among soft scalar masses.
It is not difficult to make $\tilde m_{H_1}^2$ different from 
$\tilde m_{H_2}^2$ keeping their positivity. In fact, if we take 
$\Theta=(1/\sqrt 6, -\sqrt 2/3, -\sqrt{11}/6, -\sqrt{11}/6)$,
we obtain a non-universal example such as 
\begin{eqnarray}
&&\tilde m_{\hat I}^2={5\over 24}m_{3/2}^2, \quad
\tilde m_{H_1}^2={1\over 3}m_{3/2}^2, \quad
\tilde m_{H_2}^2={1\over 12}m_{3/2}^2, \quad
\tilde m_S^2={1\over 12}m_{3/2}^2, \nonumber \\
&&\qquad M_a={1\over \sqrt 2}m_{3/2},\quad
{A_{\hat I\hat JH_\alpha}\over h_{\hat I\hat JH_\alpha}}
={A_\lambda\over\lambda}={A_k\over k}=-{1\over \sqrt 2}m_{3/2}.
\label{eqaf}
\end{eqnarray}
Now the first condition of eq.~(\ref{eqaa}) is rewritten as 
$\Theta_3^2>1/3$.
As an simple way to satisfy this condition, 
we shift the above $\Theta$ into
$\Theta=(1/3, -\sqrt 2/3, -\sqrt 2/3, -2/3)$. 
This choice keeps the soft scalar masses 
the same as in eq.~(\ref{eqaff}) except for $\tilde m_S^2$
and they are expressed as
\begin{eqnarray}
&&\tilde m_{\hat I}^2=\tilde m_{H_\alpha}^2=-\tilde m_S^2={1\over 3}m_{3/2}^2, 
\qquad M_a={1\over \sqrt 3}m_{3/2},\nonumber \\
&&{A_{\hat I\hat JH_\alpha}\over h_{\hat I\hat JH_\alpha}}=
-{\sqrt 2-1\over \sqrt 3}m_{3/2}, \qquad
{A_\lambda\over\lambda}={A_k\over k}=-{1\over \sqrt 3}m_{3/2}.
\label{eqag}
\end{eqnarray}
Note that with the above chosen $\Theta$'s which led to eqs.~(31) and (32), the
sum rule (25) is partially satisfied according to the discussion in 
sect. 2. In fact, in eq.~(32) the sum rule is broken in the down sector
and in eq.~(33) it is broken in the $\lambda$-term.
We can easily introduce more non-universality among these parameters by
changing $\Theta$ from this.
 
Here we presented simple examples. However,
as it is clearly seen from these examples, we can find that it is not 
difficult to prepare the set of parameters in a way that 
our scenario could be potentially realized.
Athough all soft parameters are generally of the same order in the present
scheme, even within these examples we find that rather extensive and 
phenomenologically allowable non-universal soft supersymmetry breaking
structure can be realized.
It is useful to note that their hierarchical structure can 
be also realized if we take into account the one-loop effect \cite{ksyy}.
Moreover, if we extend our consideration to a set of modular weights which 
have not been known to be realized in the concrete superstring models, 
we can construct many models with different patterns of the 
soft breaking parameters. 
We can also expect that this scenario might be
straightfowardly extended to the soft supersymmetry breaking parameters 
in the M-theory framework.\\

\noindent
{\large\bf 4.~Reduction of couplings and soft scalar masses sum rule}

The above proposed scenario can also be realised in a simpler field
theoretical framework of supersymmetric GUTs in which {\it reduction of
couplings} has been achieved. The method of reducing the couplings
consists of hunting for renormalization group invariant (RGI) relations.
This programme, called Gauge--Yukawa unification scheme, applied in the
dimensionless couplings of supersymmetric GUTs, such as gauge and
Yukawa couplings, had already noticeable successes by predicting
correctly, among others, the top quark mass in the finite and in the
minimal N=1 supersymmetric SU(5) GUTs \cite{finite1,kmz1}.  
An impressive aspect of the
RGI relations is that one can guarantee their validity to all-orders
in perturbation theory by studying the uniqueness of the resulting
relations at one-loop, as was proven in the early days of the
programme of reduction of couplings \cite{zim1}. 
Even more remarkable
is the fact that it is possible to find RGI relations among couplings
that guarantee finiteness to all-orders in perturbation
theory \cite{LPS,ermushev1}. 

Let us  outline briefly the idea of reduction of couplings.  
Any RGI relation among couplings 
(which does not depend on the renormalization
scale $\mu$ explicitly) can be expressed,
in the implicit form $\Phi (g_1,\cdots,g_A) ~=~\mbox{const.}$,
which
has to satisfy the partial differential equation (PDE)
\bea
\mu\,\frac{d \Phi}{d \mu} &=& {\vec \nabla}\cdot {\vec \beta} ~=~ 
\sum_{a=1}^{A} 
\,\beta_{a}\,\frac{\partial \Phi}{\partial g_{a}}~=~0~,
\eea
where $\beta_a$ is the $\beta$-function of $g_a$.
This PDE is equivalent
to a set of ordinary differential equations, 
the so-called reduction equations (REs) \cite{zim1},
\be
\beta_{g} \,\frac{d g_{a}}{d g} =\beta_{a}~,~a=1,\cdots,A~,
\label{redeq}
\ee
where $g$ and $\beta_{g}$ are the primary 
coupling and its $\beta$-function,
and the counting on $a$ does not include $g$.
Since maximally ($A-1$) independent 
RGI ``constraints'' 
in the $A$-dimensional space of couplings
can be imposed by the $\Phi_a$'s, one could in principle
express all the couplings in terms of 
a single coupling $g$.
 The strongest requirement is to demand
 power series solutions to the REs,
\be
g_{a} = \sum_{n=0}\rho_{a}^{(n)}\,g^{2n+1}~,
\label{powerser}
\ee
which formally preserve perturbative renormalizability.
Remarkably, the 
uniqueness of such power series solutions
can be decided already at the one-loop level \cite{zim1}.

The method of reducing the dimensionless couplings has been
extended \cite{kmz2}
to the soft
  supersymmetry breaking (SSB) dimensionful parameters of $N = 1$
  supersymmetric theories.   In addition it was found \cite{kkk1} that
  RGI SSB scalar masses in Gauge-Yukawa unified models satisfy a
  universal sum rule.  
Let us briefly describe here how the use of the
  available two-loop RG functions and the requirement of finiteness of
  the SSB parameters up to this order leads to the soft scalar-mass
  sum rule \cite{kkmz1}.

Consider the superpotential given by\footnote{In this section we change
 the notation for the Yukawa couplings and the soft supersymmetry
 breaking parameters from the previous section as follows:
$h_{IJK}\rightarrow C_{ijk}$, ~$A_{IJK}\rightarrow h_{ijk}$,
~$\tilde m_{I\bar J}^2 \rightarrow (m^2)_i^j$.} 
 \bea
W&=& \frac{1}{2}\,m_{ij} \,\Phi_{i}\,\Phi_{j}+
\frac{1}{6}\,C_{ijk} \,\Phi_{i}\,\Phi_{j}\,\Phi_{k}~,
\label{supot}
\eea
where $m_{ij}$ and $C_{ijk}$ are gauge invariant tensors and
the matter field $\Phi_{i}$ transforms
according to the irreducible representation  $R_{i}$
of the gauge group $G$. 
The Lagrangian for SSB terms along this superpotential is
\bea
-{\cal L}_{\rm SB} &=&
\frac{1}{6} \,h^{ijk}\,\phi_i \phi_j \phi_k
+
\frac{1}{2} \,b^{ij}\,\phi_i \phi_j
+
\frac{1}{2} \,(m^2)^{j}_{i}\,\phi^{*\,i} \phi_j+
\frac{1}{2} \,M\,\lambda \lambda+\mbox{H.c.},
\eea
where the $\phi_i$ are the
scalar parts of the chiral superfields $\Phi_i$ , 
$\lambda$ are the gauginos
and $M$ their unified mass.
Considering first finite theories, we assume that 
the gauge group is  a simple group and the one-loop
$\beta$-function of the 
gauge coupling $g$  vanishes.
We also assume that the reduction equations 
admit power series solutions of the form
\bea 
C^{ijk} &=& g\,\sum_{n=0}\,\rho^{ijk}_{(n)} g^{2n}~.
\label{Yg}
\eea 
According to the finiteness theorem \cite{LPS}, the theory is then
finite to all-orders in 
perturbation theory, if, among others, the one-loop anomalous dimensions
$\gamma_{i}^{j(1)}$ vanish.  The one- and two-loop finiteness for
$h^{ijk}$ can be achieved by 
\bea h^{ijk} &=& -M C^{ijk}+\dots =-M
\rho^{ijk}_{(0)}\,g+O(g^5)~.
\label{hY}
\eea

Now, to obtain the two-loop sum rule for 
soft scalar masses which guarantees together with condition 
(40) the finiteness of the theory in its SSB sector, we assume that 
the lowest order coefficients $\rho^{ijk}_{(0)}$ 
and also $(m^2)^{i}_{j}$ satisfy the diagonality relations
\bea
\rho_{ipq(0)}\rho^{jpq}_{(0)} &\propto & \delta_{i}^{j}~\mbox{for all} 
~p ~\mbox{and}~q~~\mbox{and}~~
(m^2)^{i}_{j}= m^{2}_{j}\delta^{i}_{j}~,
\label{cond1}
\eea
respectively.
Then we find the following soft scalar-mass sum
rule
\bea
(~m_{i}^{2}+m_{j}^{2}+m_{k}^{2}~)/
M M^{\dag} &=&
1+\frac{g^2}{16 \pi^2}\,\Delta^{(1)}+O(g^4)~
\label{sumr} 
\eea
for i, j, k with $\rho^{ijk}_{(0)} \neq 0$, where $\Delta^{(1)}$ is
the two-loop correction
\bea
\Delta^{(1)} &=&  -2\sum_{l} [(m^{2}_{l}/M M^{\dag})-(1/3)]~T(R_l),
\label{delta}
\eea
which vanishes for the
universal choice in accordance with previous findings \cite{jack1}.

If we know higher-loop $\beta$-functions explicitly, we can follow the same 
procedure and find higher-loop RGI relations among SSB terms.
However, the $\beta$-functions of the soft scalar masses are explicitly
known only up to two loops.
In order to obtain higher-loop results, we need something else instead of 
knowledge of explicit $\beta$-functions, e.g. some relations among 
$\beta$-functions.

The recent progress made using the spurion technique
\cite{delbourgo1,girardello1} leads to
the following  all-loop relations among SSB
$\beta$-functions \cite{hisano1}$-$\cite{jack4}. 
\bea
\beta_M &=& 2{\cal O}\left({\beta_g\over g}\right)~,
\label{betaM}\\
\beta_h^{ijk}&=&\gamma^i{}_lh^{ljk}+\gamma^j{}_lh^{ilk}
+\gamma^k{}_lh^{ijl}\nn\\
&&-2\gamma_1^i{}_lC^{ljk}
-2\gamma_1^j{}_lC^{ilk}-2\gamma_1^k{}_lC^{ijl}~, \label{betah}\\
(\beta_{m^2})^i{}_j &=&\left[ \Delta 
+ X \frac{\partial}{\partial g}\right]\gamma^i{}_j~,
\label{betam2}\\
{\cal O} &=&\left(Mg^2{\partial\over{\partial g^2}}
-h^{lmn}{\partial
\over{\partial C^{lmn}}}\right)~,
\label{diffo}\\
\Delta &=& 2{\cal O}{\cal O}^* +2|M|^2 g^2{\partial
\over{\partial g^2}} +\tilde{C}_{lmn}
{\partial\over{\partial
C_{lmn}}} +\tilde{C}^{lmn}{\partial\over{\partial C^{lmn}}}~,
\eea
where $(\gamma_1)^i{}_j={\cal O}\gamma^i{}_j$, 
$C_{lmn} = (C^{lmn})^*$, and 
\bea
\tilde{C}^{ijk}&=&
(m^2)^i{}_lC^{ljk}+(m^2)^j{}_lC^{ilk}+(m^2)^k{}_lC^{ijl}~. \label{tildeC}
\eea
The X in eq.~(46) in the lowest order is 
\begin{equation}
X^{(2)}=-{S g^3\over 8\pi^2}, \quad 
S\delta_{AB}=(m^2)^k_l(R_AR_B)^l_k-\vert M\vert^2C(G)\delta_{AB}.
\end{equation} 
Then assuming more generally than the finite case, described above,
(a) the existence of a RGI surface on which $C = C(g)$ or
equivalently that
\be
 \frac{dC^{ijk}}{dg} = \frac{\beta^{ijk}_C}{\beta_g}  \label{Cbeta}
\ee
holds, and (b) the existence of RGI surface on which equation
\bea
h^{ijk} &=& -M (C^{ijk})'
\equiv -M \frac{d C^{ijk}(g)}{d \ln g}~,
\label{h2}
\eea
holds too and using the all-loop gauge
$\beta$-function of Novikov {\em et al.} \cite{novikov1} given
by 
\bea
\beta_g^{\rm NSVZ} &=& 
\frac{g^3}{16\pi^2} 
\left[ \frac{\sum_l T(R_l)(1-\gamma_l /2)
-3 C(G)}{ 1-g^2C(G)/8\pi^2}\right]~, 
\label{bnsvz}
\eea 
it was found the all-loop RGI sum rule \cite{kkz},
\bea
m^2_i+m^2_j+m^2_k &=&
|M|^2 \{~
\frac{1}{1-g^2 C(G)/(8\pi^2)}\frac{d \ln C^{ijk}}{d \ln g}
+\frac{1}{2}\frac{d^2 \ln C^{ijk}}{d (\ln g)^2}~\}\nn\\
& &+\sum_l
\frac{m^2_l T(R_l)}{C(G)-8\pi^2/g^2}
\frac{d \ln C^{ijk}}{d \ln g}~.
\label{sum2}
\eea

The lowest order relations take very simple form and we may summarize
them as\footnote{Here we neglect the new $CP$ violating phases in the
soft breaking parameters.}
\begin{eqnarray}
&&m_i^2+m_j^2+m_k^2=M^2,   
\label{lowsum}\\
&&\qquad h^{ijk}=-MC^{ijk}.
\end{eqnarray}
It is interesting enough to note that these formulae have the same form 
as eq.~(\ref{eqz}). 
These formulae can be used directly
in applications using GUTs in which has been assumed reduction of coupling
beyond the unification point \cite{kkmz1}.
\\

\noindent
{\large\bf 5.~Phenomenological features}

In the previous sections it was shown that we could construct the models
which potentially induced the $\mu$-scale at the 
high energy region such as the 
unification scale. In this section we study the lightest neutral Higgs 
scalar mass bound as one of their phenomenological
features. 
Before proceeding this study we start our discussion with the 
vacuum instability problem 
due to the non-renormalizable terms in the Langangian. 
It has been known that there appears a vacuum instability
in the models with a singlet scalar chiral superfield if we take account 
of the gravity effects \cite{nonren}. 
In such models it has been shown, taking the cutoff scale $\Lambda$ to 
be of order $M_{\rm pl}$, that the dangerous divergent diagram is 
propotional to 
$M_{\rm pl}^{2-E_c-P_c}$ where $E_c$ stands for the number of external 
lines and $P_c$ represents the number of chiral propagators.
If there is no symmetry to prohibit the tadpole, then 
the tadpole contribution with $E_c=1$ violates the weak scale stability.
Although in our scenario the chiral superfield $S$ is an SM singlet
in the case of the model with an extra U(1), 
it has a non-trivial charge under the U(1).
The divergence is at most logarithmic in the case 
of $E_c=2$ and the vacuum instability seems to be escapable due to the 
extra U(1)-symmetry. This situation does not change even if this
symmetry breaks down, as in our model, at $M_{\rm pl}$.
In the NMSSM with a discrete symmetry the same situation can be expected.

The electroweak radiative symmetry breaking can also 
be one of the crucial check
points for the validity of the proposed scenario. 
Although the $\mu$-scale is induced at the 
unification scale in the present scheme, the models cannot be reduced 
to the MSSM at the unification scale. Since the vacuum expectation value of 
$\tilde S$ is comparable to the weak scale, 
this chiral superfield cannot be integrated out 
to replace the Yukawa coupling 
$\lambda SH_1H_2$ by an effective $\mu$-term at such a high energy scale. 
This singlet chiral superfield $S$ does not decouple until near the 
weak scale. Thus we should treat our models in the same way as the
usual ones with $\tilde m_S^2>0$. 
However, it is expected that there is a large
difference in the allowed parameter space  
between the models with a different sign of $\tilde m_S^2$ since the
couplings
behavior required to realize the correct vacuum radiatively is not the same. 
This difference can be reflected in 
the mass bound of the lightest neutral Higgs scalar, for example.
In the remaining part of this section 
we estimate the lightest neutral Higgs mass bound 
in the present scenario by imposing the condition of the 
weak scale radiative symmetry breaking. We compare 
these results with the ones of 
other models with $\tilde m_S^2>0$ where the $\mu$-scale 
appears around the weak scale.

At first we discuss the model with an extra U(1). 
In this model the tree level scalar potential 
including the soft supersymmetry breaking terms is the same as the usual 
$\tilde m_S^2>0$ case and can be written as\footnote{
Hereafter we will use the same notation for the superfield and its
scalar component. The sign convention of $A$ parameters is followed to
the one in the previous sections which is reversed from that in
Refs. \cite{ds,ds2}. }
\begin{eqnarray}
V_0&=&{1\over 8}\left(g_2^2+g_1^2\right)\left(
\vert H_1\vert^2-\vert H_2\vert^2\right)^2 
+\left(\vert\lambda SH_1\vert^2+\vert\lambda SH_2\vert^2\right)\nonumber\\
&+&\tilde m_{H_1}^2\vert H_1\vert^2+\tilde m_{H_2}^2\vert H_2\vert^2 
+(A_\lambda\lambda
SH_1H_2+{\rm h.c.})\nonumber\\
&+&{1\over 8}g_E^2\left(Q_1\vert H_1\vert^2
+Q_2\vert H_2\vert^2+Q_S\vert S\vert^2\right)^2 
+\lambda^2\vert H_1H_2\vert^2 +\tilde m_S^2\vert S\vert^2,
\label{eqah}
\end{eqnarray}
where $Q_1$, $Q_2$ and $Q_S$ are the extra U(1)-charges of $H_1$, $H_2$ 
and $S$, respectively.
The vacuum of these models is parametrized by the vacuum expectation 
values (VEVs) of Higgs scalar fields such as
\begin{equation}
\langle H_1\rangle=\left(\begin{array}{c}v_1 \\0\\ \end{array}\right),
\qquad \langle H_2\rangle=\left(\begin{array}{c}0 \\v_2\\ 
\end{array}\right),
\qquad
\langle S\rangle=u,
\label{eqai}
\end{equation}
where $v_1$ and $v_2$ are assumed to be positive and $v_1^2+v_2^2
=v^2(\equiv (174{\rm GeV})^2)$ should be satisfied. 
Thus the vacuum in these models is parametrized by $\tan\beta=v_2/v_1$ 
and $u$.
Based on this scalar potential the upper mass bound of the lightest
neutral Higgs scalar is estimated as
\begin{equation}
m_{h^0}^{(0)2} \le m_Z^2\left[\cos^22\beta +{2\lambda^2\over
g_1^2+g_2^2}\sin^22\beta+{g_E^2 \over
g_1^2+g_2^2}\left(Q_1\cos^2\beta+Q_2\sin^2\beta\right)^2\right]. 
\label{eqaj}
\end{equation}
In this derivation we have used the following potential minimization 
conditions for $V_0$ :
\begin{eqnarray}
\hspace*{-7mm}&&\tilde m_{H_1}^2=-{1\over 4}(g_2^2+g_1^2)(v_1^2-v_2^2)-{1\over
4}g_E^2Q_1(Q_1v_1^2+Q_2v_2^2+Q_Su^2) -\lambda^2(u^2+v_2^2)
-\lambda A_\lambda u{v_2\over v_1}, \nonumber \\
\hspace*{-7mm}&&\tilde m_{H_2}^2={1\over 4}(g_2^2+g_1^2)(v_1^2-v_2^2)-{1\over
4}g_E^2Q_2(Q_1v_1^2+Q_2v_2^2+Q_Su^2) -\lambda^2(u^2+v_1^2)
-\lambda A_\lambda u{v_1\over v_2}, \nonumber \\
\hspace*{-7mm}&&\tilde m_S^2=-{1\over 4}g_E^2Q_S(Q_1v_1^2+Q_2v_2^2+Q_Su^2)
-\lambda^2(v_1^2+v_2^2)- \lambda A_\lambda{v_1v_2\over u}.
\label{eqak}
\end{eqnarray}
These formulas can be extended so as to include the one-loop effective
potential
\begin{equation}
V_1={1 \over 64\pi^2}{\rm Str}~ 
{\cal M}^4~ \left(\ln{{\cal M}^2 \over \Lambda^2}-{3\over 2}\right),
\label{eqal}
\end{equation}  
where ${\cal M}^2$ is a matrix of the squared mass of the fields
contributing to the one-loop correction and $\Lambda$ is a 
renormalization point. 
If we include the effect of $V_1$,
the upper mass bound of the lightest Higgs scalar is modified by 
adding the following term to the right-hand side of eq.~(\ref{eqaj}) : 
\begin{equation}
\Delta m_{h^0}^2={1\over 2}\left({\partial^2 V_1 \over\partial
v_1^2}-{1\over v_1}{\partial V_1\over \partial v_1}\right)\cos^2\beta
+{1\over 2}{\partial^2 V_1 \over\partial v_1\partial v_2}\sin 2\beta
+{1\over 2}\left({\partial^2 V_1 \over\partial v_2^2}
-{1\over v_2}{\partial V_1\over \partial v_2}\right)\sin^2\beta.  
\label{eqam}
\end{equation}
This bound is estimated at the minimum of the one-loop effective potential 
$V_{\rm eff}=V_0+V_1$. 

In order to proceed this estimation we will use the
following method which was adopted in Ref. \cite{ds}. 
As usual, we evolve the coupling constants and 
the soft supersymmetry breaking parameters from the suitable initial values 
at the unification scale $M_X$ towards the low energy region 
by using the renormalization group equations (RGEs).
We can determine $\tan\beta$ using the data on the top quark mass $m_t$ and 
the top Yukawa coupling constant $h_t(m_t)$\footnote{To determine 
the value of $\tan\beta$ an effect of
the translation of the running mass to the pole mass has been 
taken into account \cite{pole}.}.
Once $\tan\beta$ is fixed, we can determine the value of $u$ 
which minimizes the effective potential $V_{\rm eff}$. 
However, this set of $(\tan\beta,u)$ is not necessarily a true potential 
minimum. In order to guarantee that these values are indeed obtained
at the potential minimum we substitute these values of $(\tan\beta,u)$ 
into potential minimum conditions (\ref{eqak}) improved by $V_1$ 
and compare them with the corresponding direct low energy results of the 
RGEs. 
If these are equal within the suitable range, we recognize such
$(\tan\beta,u)$ as the true vacuum and the corresponding parameters set 
as the phenomenologically allowable one.
Only for such parameters set we estimate the upper mass bound of the
lightest Higgs scalar $m_{h^0}$ which is defined as
$m_{h^0}^2=m_{h^0}^{(0)2}+\Delta m_{h^0}^2$. 
In this calculation we change the RGEs from the
ones of the supersymmetric model to the ones of the nonsupersymmetric model at
$m_{3/2}$. The renormalization point in eq.~(\ref{eqal}) is taken as $m_t$, for
simplicity.
Although these are the main points of our analysis, we should mention that
we additionally impose further phenomenological conditions 
such as the mass bounds of superparticles to restrict the parameters, 
which have also been used in the analysis of \cite{ds}. 

In our analysis we have used first eqs. (\ref{eqaff})~ and~ (\ref{eqag})~ for 
the initial values of the soft supersymmetry breaking parameters
as representative cases of~ $\tilde m_S^2>0$ and $\tilde m_S^2<0$,
respectively. However,
it seems to be rather difficult to obtain consistent radiative
symmetry breaking solutions for the universal soft Higgs masses in the
case of $\tilde m_S^2>0$.
Therefore, we are led to use eq. (\ref{eqaf}) instead of
eq. (\ref{eqaff})
in order to represent the $\tilde m_S^2>0$ case. On the other hand 
as a typical model with an extra U(1) we take the three generations 
$\xi_-$ model studied in Ref. \cite{ds} which is derived from $E_6$
using the Wilson line breaking mechanism.
The SM gauge coupling unification is guaranteed for the field contents
of this model since the chiral superfield contents are composed of the
MSSM one with $3({\bf 5}+\bar{\bf 5})$ of SU(5). 
The relevant Yukawa couplings to the
RGEs analysis are the top Yukawa coupling $h_t$, $\lambda$ and $k$\footnote{
We may consider the model with no extra colored fields $(g,\bar g)$
in the $\tilde m_S^2<0$ case as an interesting alternative.
We do not consider such a case here since further consideration
on the matter content for the anomaly cancellation is required.}. 
They are surveyed in the range from 0 to 3 with the interval 0.1, 
respectively. 
The results of this analysis for $m_{3/2}=1$~TeV are summarized in Fig.1.

\input epsf 
\begin{figure}[tb]
\begin{center}
\epsfxsize=7.6cm
\leavevmode
\epsfbox{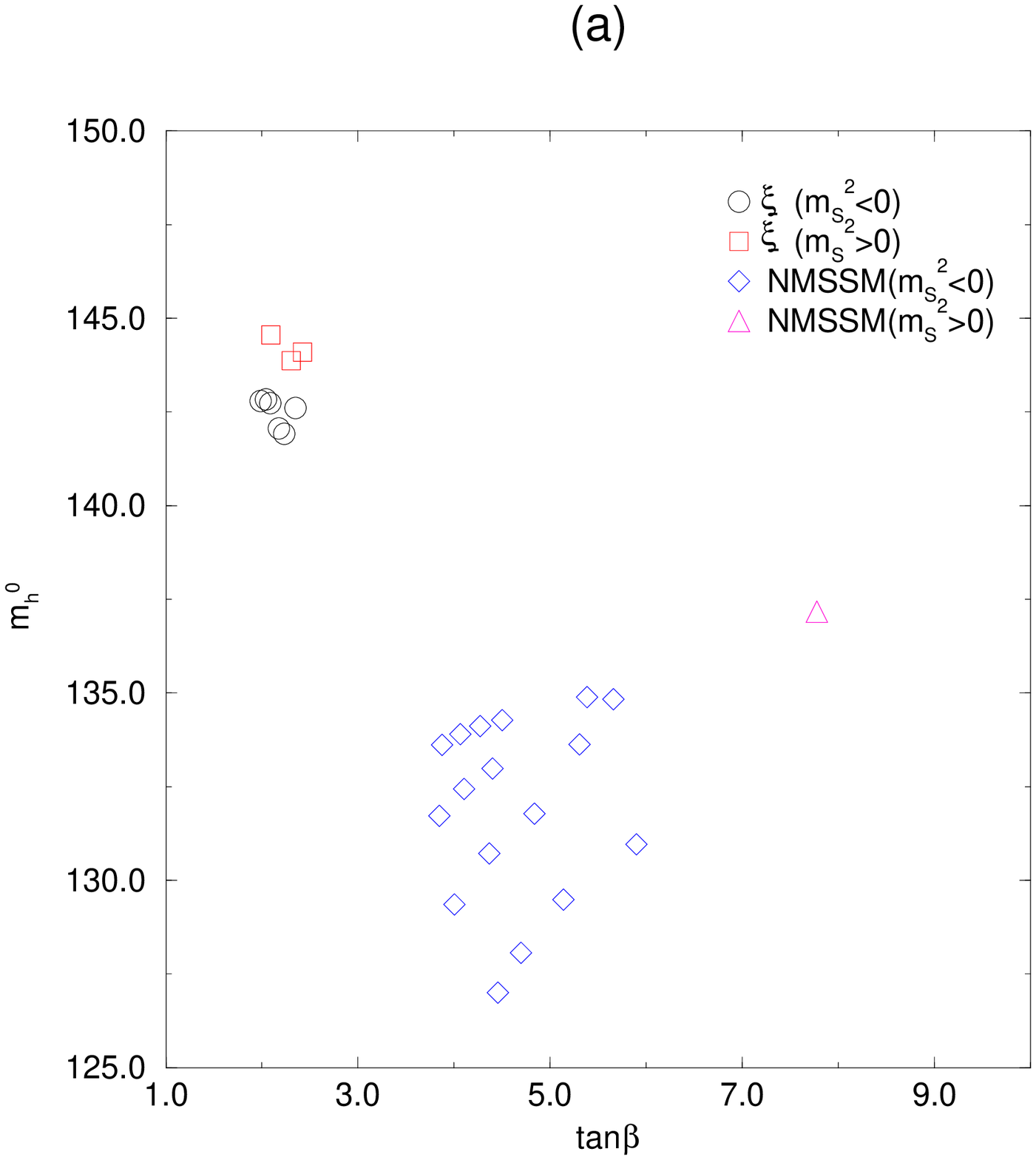}
\hspace*{0.3cm}
\epsfxsize=7.6cm
\leavevmode
\epsfbox{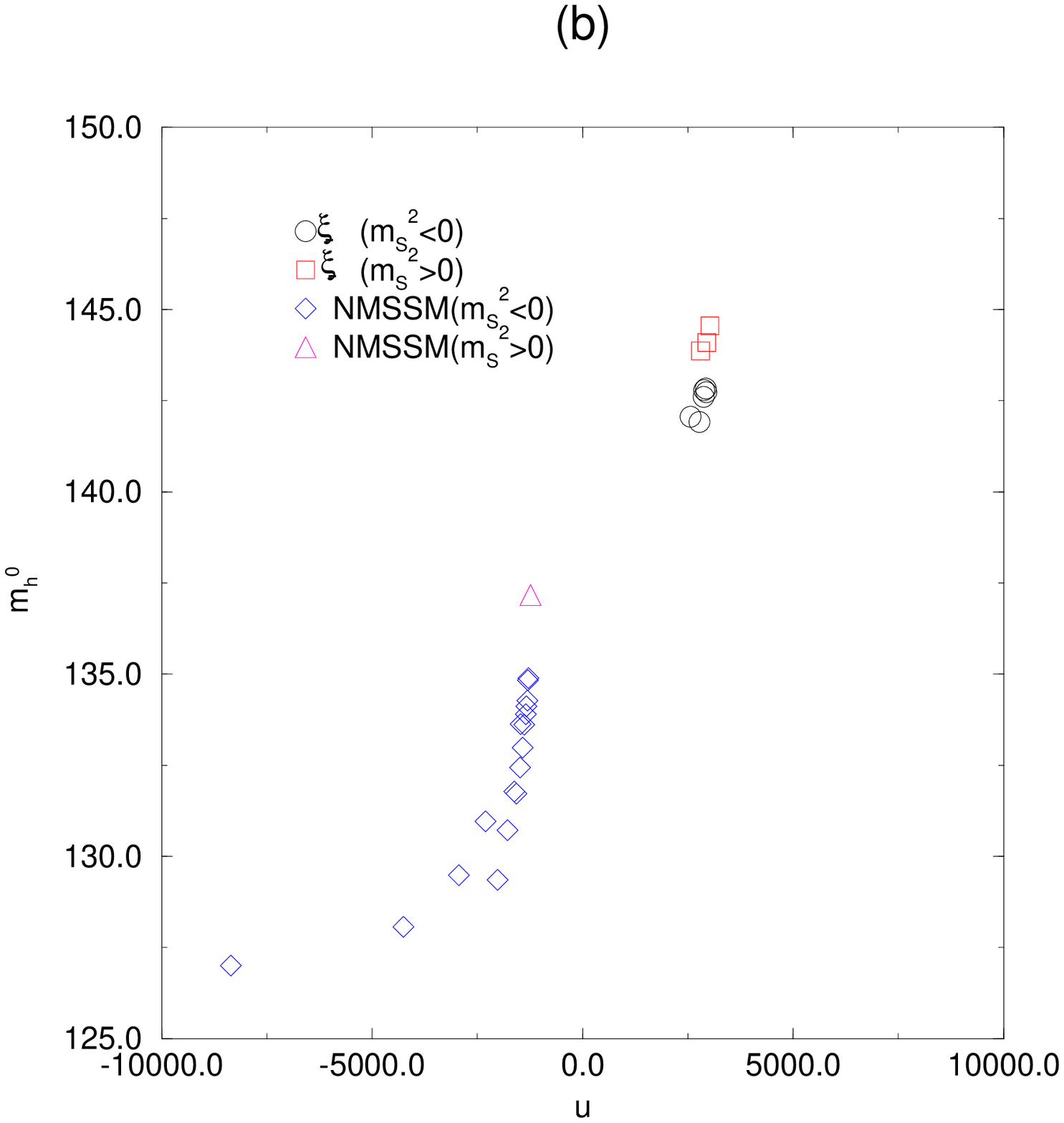}
\end{center}
\vspace*{-1cm}
{\footnotesize Fig.1~~\  Scatter plots of the upper mass bound
 $m_{h^0}$ of the
 lightest neutral Higgs scalar for $\tan\beta$ in Fig. 1(a) and for $u$ 
in Fig. 1(b) for both of the $\xi_-$ model and the NMSSM. 
They are plotted in both cases of
$\tilde m_S^2>0$ and $\tilde m_S^2<0$.}
\end{figure}

The allowed parameter space seems to be strongly restricted in the case of
$\tilde m_S^2>0$ so that the number of solutions is limited as compared 
to the one of $\tilde m_S^2<0$. 
The sign of $\tilde m_S^2$ at the unification scale affects largely 
the radiative symmetry breaking phenomena.
It seems to bring different features
for the running of the coupling constants in both cases, although there
is no large difference in their initial values at the unification scale.
The consistent radiative symmetry breaking tends to occur much more
easily in the case of $\tilde m_S^2<0$. 
The upper mass bound $m_{h^0}$ takes its value at $\sim$142~GeV~
($\tilde m_S^2<0$ case) and
$\sim$144~GeV~($\tilde m_S^2>0$ case).
We might expect that the $\tilde m_S^2<0$ case  
generally gives smaller values for the $m_{h^0}$ as compared to 
the $\tilde m_S^2>0$ one.
We also carried out the calculation for other values of $m_{3/2}$.
The qualitative feature is very similar, although the numerical value
of $m_{h^0}$ tends to be shifted upwards for the larger value of $m_{3/2}$.
It is also useful to note that in the model with an extra U(1) 
values of $\tan\beta$ smaller than 2 are allowed even for the present Higgs 
mass bound
since there is the extra contribution depending on $\lambda$ as shown in 
eq.~(\ref{eqaj})\footnote{The results shown in Fig.1 are
obtained for rather restricted soft parameters and then this aspect can
be not seen explicitly. This feature of allowed $\tan\beta$
will be expected in the NMSSM \cite{ds,ds2}.}.

Next we proceed to the analysis of the NMSSM.
The NMSSM is another candidate to solve the $\mu$-problem in the MSSM.
As mentioned in the introduction, the discrete symmetry should be
imposed in the NMSSM to escape the phenomenological problems
such as the bare $\mu$-term and the tadpole.
Although, in general, the spontaneous breaking of the discrete 
symmetry generates the cosmologically dangerous domain wall problem,
we leave it here.
We estimate the upper mass bound of the
Higgs scalar also in the NMSSM paying attention to the difference between
the $\tilde m_S^2<0$ and $\tilde m_S^2>0$ cases. 
The NMSSM is composed of the MSSM content and a singlet chiral 
superfield. 
Here we do not consider the introduction of the extra matter fields such 
as $5+\bar 5$ of SU(5)\footnote{The inclusion of these is expected to
raise the Higgs upper mass bound as discussed in \cite{ds2}.}.
In the superpotential we should 
replace $k Sg\bar g$ in the $\xi_-$ model to $ {k\over 3} S^3$.
As the supersymmetry breaking parameters in this study, we use 
those given in eq.~(\ref{eqaf}) with 
$A_k/k=(\sqrt{33}+3\sqrt 2-2\sqrt 6)m_{3/2}/6$ and the ones in
eq.~(\ref{eqag}) with $A_k/k=(5-2\sqrt 2)m_{3/2}/\sqrt 3$ 
for the cases of $\tilde m_S^2>0$ and $\tilde m_S^2<0$, respectively. 
The Yukawa couplings $h_t$, $\vert k\vert$ and $\lambda$ are serveyed 
in the range from 0 to 3 with an interval 0.1, while
$m_{3/2}$ is fixed to be 1~TeV.

The results of this analysis are also presented in Fig.1.
The allowed parameter space again seems to be strongly 
restricted in the $\tilde
m_S^2>0$ case as compared to the $\tilde m_S^2<0$ case.
Since there is no direct contribution to the RGE of 
$\tilde m_S^2$ from the Yukawa
coupling which is affected by the SU(3) gauge coupling in the NMSSM,
it is rather difficult to make $\tilde m_S^2$ negative radiatively and 
this tendency becomes much clearer as compared to the model with an 
extra U(1).
The upper mass bound $m_{h^0}$ takes its value in the region
of 127~GeV$\sim$135~GeV ~($\tilde m_S^2<0$ case) and $\sim$137~GeV
~($\tilde m_S^2>0$ case). 
Also in the NMSSM the value of $m_{h^0}$ is smaller in the 
$\tilde m_S^2 <0 $ case. 
In this case the smaller value of $m_{h^0}$ tends to be realized
at the large $\vert u\vert$ region where there is no solutions 
in the $\tilde m_S^2>0$ case. 
Since only the effect of $k$ and $\lambda$ cannot make 
the value of $\tilde m_S^2$ negative
largely, the negative initial value of $\tilde m_S^2$ seems to be
necessary to derive the larger $\vert u\vert$ value in the NMSSM. 
The large $\vert u\vert$ value tends to be combined with 
the smaller $\lambda$ since $\lambda u$ is related to the $\mu$-scale.   
This feature suggests that the $\tilde m_S^2<0$ model can have the possibility 
to realize a smaller upper mass bound of $m_{h^0}$.

Up to now we focussed our attention on the soft supersymmetry breaking
parameters which could be derived in the superstring framework.
As we have discussed already, the soft supersymmetry breaking scalar masses
appear also in the field theoretical framework and the corresponding sum
rule (\ref{lowsum}) can be obtained there too assuming the 
reduction of couplings
scenario. In the latter framework we shall study the effect of 
$\tilde m^2_S$ on the Higgs mass bound under the constraint 
coming from the sum rule. We should note that
there are some noticeable points which are characteristic in this study. 
The Gauge-Yukawa and Finite SU(5) models [23,24], which have been 
so far successful in predicting, among others, the top quark mass require large
values of $\tan\beta$ since the ratio of the top and bottom Yukawa couplings
$h_t$ and $h_b$ (as well their ratio to the GUT gauge coupling, 
$g_{\rm GUT}$) are of order one at the unification scale. 
This situation is completely different from the above superstring case, 
where the sum rule is required only for the Yukawa
couplings which satisfy a condition such as (\ref{eqy}). 
However the field theory framework, according to the partial 
reduction of couplings proposal [41], is flexible on the choice 
of the couplings that should be reduced. Therefore one can 
always reduce the number of the sum rules that have to be satisfied.
Therefore, as a first try, assuming that a successful GUT could
be constructed with reduction of couplings which keeps the above
characteristics of the known Gauge-Yukawa GUTs and moreover incorporates a
term such as $\lambda SH_1H_2$ with a coupling which is reduced in favor of
the corresponding $g_{\rm GUT}$, we shall study the consequences of 
the sum rule applied in this term\footnote{Note that a singlet cannot 
have non-vanishing couplings in Finite N = 1 GUTs, resulting from 
the finiteness condition that requires vanishing of the corresponding 
anomalous dimension of the chiral singlet field. 
This is not the case in other Gauge-Yukawa GUTs.}. 
Then the relation $\tilde m_{H_1}^2+\tilde m_{H_2}^2+\tilde m_S^2=M^2$ 
should be
satisfied and clearly the value of $\tilde m_S^2$ reflects to those 
of $\tilde m_{H_1}^2$
and $\tilde m_{H_2}^2$ which are already constrained from the large
$\tan\beta$. 
We shall see in the following that, as a result of imposing these
constraints, it becomes completely non-trivial to achieve 
radiative electroweak symmetry breaking.

Under the assumptions described above we take the NMSSM
within the reduction of couplings scenario as the target of our numerical
study. Then in the present RGE study we need to include the bottom
and tau Yukawa couplings $h_b,~h_{\tau}$ since they are
comparable to $h_t$ and we assume the validity of the sum rule corresponding
to all Yukawa couplings of the third generation, assuming reduction of the
corresponding couplings\footnote{Note that the coupling of a term 
such as ${k\over 3}S^3$ cannot be reduced in favor of the GUT gauge 
coupling, since $S$ is gauge singlet.}.
We can further impose the condition $h_b = h_{\tau}$ at the 
unification scale as resulting naturally in most
GUTs. 
Then we would like to determine which sum rules, as those suggested by
eq.~(\ref{lowsum}), can be imposed at the unification point and simultaneously 
are compatible with the low energy phenomenolological conditions. As a
numerical strategy to find solutions to this requirement we use
the sum rule condition in the form\footnote{Note that 
we use the notation of section 2 here.} of $A_{IJK}=-Mh_{IJK}$ and 
$\tilde m_I^2+\tilde m_J^2+\tilde m_K^2={\cal C}M^2$, 
in which ${\cal C}$ is treated as the free parameter
satisfying $0\le {\cal C}\le 3$\footnote{
We assume ${\cal C}_t={\cal C}_b$ which is the case for the 
grand unified model with
the reduction of couplings. For the $\tau$ Yukawa coupling, however,
${\cal C}_\tau$ is fixed as 3
and also $\tilde m_{\bar E}^2=M^2$ since the small ${\cal C}_\tau$ tends to 
violate the superparticle mass bound, in particular, the chargino mass 
bound and the electric charge conservation.
This obviously consists an additional
difficulty in finding the prescribed solutions.}.  
Moreover, we treat soft scalar masses as the independent free parameters
and vary $\vert \tilde m_I\vert$ from $0.1M$ to $M$,
although the gaugino masses are treated as the universal value $M$.
We impose the constraint of consistent 
realization of the bottom and tau mass at the low energy scale 
in addition to the previously explained procedure.
We take $m_{3/2}=1$~TeV as the scale where we change the set of RGEs
and the gaugino mass $M$ is varied from $0.7m_{3/2}$ to $m_{3/2}$.
Yukawa couplings are serveyed in the similar region to the previous 
superstring case.

\begin{figure}[tb]
\begin{center}
\begin{tabular}{c|ccccc}\hline
         & $\tan\beta$  &  $u$~(TeV)  & $m_{h^0}$~(GeV) & ${\cal C}_\lambda$ 
& ${\cal C}_t$ \\ \hline
$\tilde m_S^2>0$~& ~$\sim 28.2$~ & ~$1.41\sim 1.80$~ & ~$135.6\sim 139.8$~ & 
~$0.82 \sim 2.13$~ & ~$1.0\sim 3.0$~ \\
$\tilde m_S^2<0$~& ~$\sim 40.1$~ & ~$1.74\sim 2.26$~ & ~$137.6\sim 142.6$~ & 
~$0.36 \sim 1.48$~ & ~$1.5\sim 3.0$~ \\ \hline
\end{tabular}
\vspace*{3mm}\\
{\footnotesize Table 1~~ The numerical results in the field theory case.}
\end{center}
\end{figure}

The solutions of this numerical calculation are presented in Table 1. 
We could obtain these solutions only for the very restricted values for 
the Yukawa couplings
\begin{eqnarray}
&&h_t \simeq 1.06,\qquad  h_b\simeq 0.32,\qquad k \simeq 0.3, \qquad 
\lambda\simeq 2.4, \qquad (\tilde m_S^2>0), \nonumber\\
&&h_t \simeq 1.08,\qquad  h_b\simeq 0.54,\qquad k \simeq 0.4, \qquad 
\lambda\simeq 2.1, \qquad (\tilde m_S^2<0)
\end{eqnarray}
at the unification scale.
The large $\tan\beta$ solution requires fine tuned Yukawa couplings
for a consistent electroweak radiative symmetry breaking.
These are easily realized for the non-universal soft Higgs scalar masses
$\tilde m_{H_1}^2\not= \tilde m_{H_2}^2$. 
In the $\tilde m_S^2<0$ case 
we have solutions only for $\tilde m_{H_1}^2\ge \tilde m_{H_2}^2$
although in the $\tilde m_S^2 >0$ case we can obtain solutions even for
$\tilde m_{H_1}^2 < \tilde m_{H_2}^2$.
The value of $\tan\beta$ is rather different for different sign of $\tilde
m_S^2$.  This situation is very different from the previous examples in
which $\tan\beta$ takes similar value in both cases.
This may be reflected to the upper mass bound of the neutral 
Higgs scalar which is a little bit smaller for the $m_S^2>0$ case 
as compared to the $m_S^2<0$ case.
Although we do not strictly impose the sum rule for the soft 
scalar masses at the unification scale, we can see whether it is 
satisfied in these solutions through the values of ${\cal C}$ in Table 1.
The value of ${\cal C}$ shown in Table 1 is found to reflect the sign of
$\tilde m_S^2$. Since the negative $\tilde m_S^2$ tends to make 
$\tilde m_{H_{1,2}}^2$ larger, we can expect that the solutions tend to have 
smaller ${\cal C}_\lambda$ and larger ${\cal C}_t$ as 
compared to the positive $\tilde
m_S^2 $. In fact, many solutions of the negative $\tilde m_S^2$ case
satisfy ${\cal C}_\lambda <1$ and ${\cal C}_t>1$.
Therefore our study shows that we cannot apply consistently our scenario
if we insist in reducing the Yukawa couplings of the third generation in
favor of $g_{\rm GUT}$, which would imply the validity of the corresponding sum
rules. Therefore we are led to abandon the freedom offered by the field
theory in reducing more couplings and we keep only as an interesting
viable possibility the reduction of the coupling $\lambda$ of the term 
$\lambda S H_1 H_2$, which in turn implies a sum rule as in the string case
discussed earlier. It is amusing that in the positive $\tilde m_S^2$ case
many solutions satisfy the sum rules and then 
${\cal C}_\lambda \sim {\cal C}_t \sim 1$ in rather good accuracy. 
If we consider the couplings reduction in the type of model as 
the NMSSM, the positive $\tilde m_S^2$ will be promising and
$\tan\beta\sim 30$ may be expected.\\

\noindent
{\large\bf 6.~Summary and conclusions}

We have considered the possible solution for the $\mu$-problem based on
the non-universal soft supersymmetry breaking scalar masses.
Usually models with negative soft squared mass 
are excluded since they are led to incorrect SM vacuum. 
However, if we consider the non-universality of the soft supersymmetry 
breaking parameters in the extended 
models of the MSSM seriously, we can obtain viable models for some
sets of their initial conditions. Here
we have proposed such a kind of application of the non-universal soft
supersymmetry breaking to resolve the $\mu$-problem. 

In this paper we have concretely showed the possibilities that the negative
soft scalar squared mass were constructed consistently 
in the superstring framework.
We have also discussed that in the field theory with the reduction of 
couplings the non-universal soft scalar masses could be realized
satisfying the suitable sum rules. In both cases after the extension of
the MSSM it seems not to be difficult to realize the negative squared 
soft scalar mass in the extended chiral superfield sector keeping 
its positivity in the sector of the MSSM at the unification scale.
The question is whether the radiative electroweak symmetry breaking can
occur starting from such a kind of initial conditions.
We have studied to which extent this scenario could be consistent with the
electroweak radiative symmetry breaking. Our numerical study shows that 
indeed the realization of such a scenario is possible, although
the solutions seem to be shifted in the parameters 
space as compared to the usual 
case with the positive squared soft scalar masses. 
As a result of this shift in the parameters space, we also suggested 
that in the superstring case the upper mass bound of the lighest 
neutral Higgs scalar could be smaller in this type of models
than the ones with the positive squared soft scalar masses.
On the other hand, in the field theory case if we insist in reducing
in favor of the gauge coupling also the Yukawa coupligs of the third
generation, in addition to the coupling of the singlet to the Higgs fields,
the sum rules for the soft supersymmetry breaking parameters can be easily
satisfied only in the case of positive soft scalar squared masses. 
Certainly a negative soft scalar squared mass corresponding to the
singlet superfield can also be obtained by relaxing the requirement to
reduce the Yukawa couplings in favor of the gauge couplings.

These results seem to be interesting enough for the extended MSSM with a SM
singlet chiral superfield.   
If we take into account the new possibility discussed here,
the extended MSSM with the SM singlet chiral superfield may reveal the
new aspect of its phenomenology and be a similarly promising candidate 
to the MSSM, for example, in the Higgs search.  
Our present scenario for the origin of the $\mu$-scale seems to be a simple
example of the phenomena that the non-universal soft supersymmetry
breaking can introduce in the theory and might exist further fruitful
applications in the model building beyond the MSSM.

\vspace*{1cm}
The work of GZ has been partially supported by the EU project 
ERBFMRXCT960090 and the A.v.Humboldt Foundation and the work of 
DS has been partly supported by the
Grant-in-Aid for Scientific Research from the Ministry of Education, 
Science and Culture(\#11640267). GZ would like to
thank the theory group (QFT) of Humboldt University in Berlin, where part
of this work was done, for their kind hospitality.
 
 \newpage
 

\begin{thebibliography}{99}
 \bibitem{n}For a review, see for example, H.-P.~Nilles, Phys. Rep.
 {\bf C110}, 1 (1984), and references therein.
 
 \bibitem{mu}J.~E.~Kim and H.~P.~Nilles, Phys. Lett. {\bf B138} (1984) 150.
 
 \bibitem{nmssm}J.~E.~Kim and H.~P.~Nilles, Phys. Lett. {\bf B263}
 (1991) 79;
 E.~J.~Chun, J.~E.~Kim and H.~P.~Nilles, Nucl. Phys. {\bf B370} (1992)
 105.
 
 \bibitem{cm}J.~A.~Casas and C.~Mu\~noz, Phys. Lett {\bf B306} (1993) 228.
 
 \bibitem{gm}G.~F.~Giudice and A.~Masiero, Phys. Lett. {\bf B206} (1988) 480.
 
 \bibitem{musol}For a review of recent works, see for example, N.~Polonsky,
 	hep-ph/9911329, and references therein.
 
 \bibitem{singlet}H.~P.~Nilles, M.~Srednicki and D.~Wyler, Phys. Lett.
 {\bf B129} (1983) 364;
 L.~E.~Ib\'a\~nez and J.~Mas, Nucl. Phys. {\bf B286} (1987) 107;
 J.~Ellis, J.~F.~Gunion, H.~E.~Haber, L.~Roszkowski and F.~Zwirner,
 Phys. Rev. {\bf D39} (1989) 844.
 
 \bibitem{color}J.-P.~Derendinger and C.~A.~Savoy, Nucl. Phys. 
 {\bf B237} (1984) 364.
 
 \bibitem{sy}D.~Suematsu and Y.~Yamagishi, Int. J. Mod. Phys. {\bf A10} 
 (1995) 4521.
 
 \bibitem{extra}M.~Cveti$\check{\rm c}$ and P.~Langacker, Phys. Rev. 
 {\bf D54} (1996) 3570.
 
 \bibitem{nonren}U.~Ellwanger, Phys. Lett. {\bf 133B} (1983) 187;
 J.~Bagger and E.~Poppitz, Phys. Rev. Lett. {\bf 71} (1993) 2380.
 
 \bibitem{cde}M.~Cveti$\check{\rm c}$, D.A.Demir, J.~R.~Espinoza, L.~Everett 
 and P.~Langacker, Phys. Rev. {\bf D56} (1997) 2861;
 P.~Langacker and J.~Wang, Phys. Rev. {\bf D58} (1998) 115010.
 
\bibitem{ds} Y.~Daikoku and D.~Suematsu, Phys. Rev. {\bf D62} (2000)
095006 (hep-ph/0003205).
 
 \bibitem{il}L.~E.~Ib\'a\~nez and D.~L\"ust, Nucl. Phys. 
 {\bf B382} (1992) 305.
 
 \bibitem{kl}V.~S.~Kaplunovsky and J.~Louis, 
 Phys. Lett. {\bf B306} (1993) 269.
 
 \bibitem{bim}A.~Brignole, L.~E~.Ib\'{a}\~{n}ez and C.~Mu\~{n}oz,
 Nucl. Phys. {\bf B422} (1994) 125.
 
 \bibitem{csgp}E.~Cremmer, S.~Ferrara, L.~Girardello and A.~Van Proeyen,
 Nucl. Phys. {\bf B212} (1983) 413.
 
 \bibitem{sum}D.~Suematsu, Phys. Rev. {\bf 54} (1996) 5715.
 
 \bibitem{ksyy}T.~Kobayashi, D.~Suematsu, K.~Yamada and Y.~Yamagishi,
 Phys. Lett. {\bf B348} (1995) 402.
 
 \bibitem{one}L.~E.~Ib\'a\~nez and H.~P.~Nilles, Phys. Lett. {\bf B169}
 (1986) 345;
 L.~Dixon, V.~Kaplunovsky and J.~Louis, Nucl. Phys. {\bf B355} (1991) 649.
 
 \bibitem{bd}Y.~Nambu, preprint EFI 92-37;
 C.~Kounnas, F.~Zwirner and I.~Pavel, Phys. Lett. {\bf B335} (1994) 403;
 P.~Binetruy and E.~Dudas, Phys. Lett. {\bf B338} (1994) 23.
   
 \bibitem{bims}A.~Brignole, L.~E.~Ib\'a\~nez, C.~Mu\~noz and
 C.~Scheich, Z. Phys. {\bf C74} (1997) 157. 

\bibitem{finite1} D. Kapetanakis, M. Mondrag\'on and G. Zoupanos,
  Zeit. f. Phys.~{\bf C60} (1993) 181; M. Mondrag\'on and G. Zoupanos,
   Nucl. Phys. {\bf B} (Proc.~Suppl.) {\bf 37C} (1995) 98.

\bibitem{kmz1}
 J. Kubo, M. Mondrag\'on and G. Zoupanos,  Nucl. Phys. {\bf B424} (1994)
291.
 
\bibitem{zim1}
 W. ~Zimmermann, Com. Math. Phys. {\bf 97} (1985) 211;
 R. ~Oehme and W. ~Zimmermann, Com. Math. Phys. {\bf 97} (1985) 569.

\bibitem{LPS} C. Lucchesi, O. Piguet and K. Sibold,
              Helv. Phys. Acta  {\bf 61} (1988) 321;
                Phys. Lett. {\bf B201} (1988) 241; 
see also C. Lucchesi and G. Zoupanos,  Fortsch. Phys. {\bf 45} (1997) 129.

\bibitem{ermushev1}
A.Z. Ermushev, D.I. Kazakov and O.V. Tarasov, Nucl. Phys. {\bf 281} (1987) 72; 
D.I. Kazakov,  Mod. Phys. Lett. {\bf A9} (1987) 663.

\bibitem{kmz2} J. Kubo, M. Mondrag\'on and G. Zoupanos,
 Phys. Lett. {\bf B389} (1996) 523. 

\bibitem{kkk1} T. Kawamura, T. Kobayashi and J. Kubo, 
Phys. Lett. {\bf B405} (1997) 64.

\bibitem{kkmz1}T.~Kobayashi, J.~Kubo, M.~Mondragon and G.~Zoupanos, 
 Nucl. Phys. {\bf B511} (1997) 45; in proc. of ICHEP 1998, vol 2, p.1597
	 (Vancouver 1988); Acta Phys. Polon. {\bf B30} (1999) 2013;
in proc. of HEP 1999, p. 804 (Tampere 1999).


\bibitem{jack1} I. Jack and D.R.T. Jones,  Phys. Lett. {\bf B333} (1994) 372.

\bibitem{delbourgo1}
 R. Delbourgo,  Nuovo Cim. {\bf 25A} (1975) 646; 
A. Salam and J. Strathdee, 
      Nucl. Phys. {\bf B86} (1975) 142; 
K. Fujikawa and W. Lang,  Nucl. Phys. {\bf B88}
     (1975) 61; M.T. Grisaru, M. Rocek and W. Siegel, 
 Nucl. Phys. {\bf B159} (1979) 429.

\bibitem{girardello1}
 L. Girardello and M.T. Grisaru, Nucl.~Phys. {\bf B194} (1982) 65; 
J.A. Helayel-Neto, Phys. Lett. {\bf B135} (1984) 78; 
 F. Feruglio, J. A. Helayel-Neto and F. Legovini,  Nucl. Phys. {\bf
	 B249} (1985) 533;
 M. Scholl,  Zeit. f. Phys. {\bf C28} (1985) 545.

\bibitem{hisano1} J. Hisano and M. Shifman,  Phys. Rev. {\bf D56} (1997) 5475.

\bibitem{jack3} I. Jack and D.R.T. Jones,  Phys. Lett. {\bf B415} (1997) 383.

\bibitem{avdeev1} L.V. Avdeev, 
D.I. Kazakov and I.N. Kondrashuk, {\em Nucl.~Phys.}~{\bf B510} (1998) 289;
D.I.Kazakov, hep-ph/9812513.

\bibitem{kazakov2}
D.I. Kazakov, M.Yu. Kalmykov, I.N. Kondrashuk
and  A.V. Gladyshev,  Nucl. Phys. {\bf B471} (1996) 387.

\bibitem{kazakov1} D.I. Kazakov,  Phys. Lett. {\bf B412} (1998) 21.

\bibitem{jack4} I. Jack, D.R.T. Jones and
 A. Pickering,  Phys. Lett. {\bf B426} (1998) 73.

\bibitem{novikov1} V. Novikov, M. Shifman, 
A. Vainstein and V. Zakharov,  Nucl. Phys. {\bf B229} (1983) 381; 
Phys. Lett. {\bf B166} (1986) 329;  
M. Shifman,  Int. J. Mod. Phys. {\bf  A11} (1996) 5761 and 
references therein.  

\bibitem{ksz}J.~Kubo, K.~Sibold and W.~Zimmermann, Nucl. Phys. {\bf B259}
	 (1985) 331.

\bibitem{kkz} T.~Kobayashi, J.~Kubo and G.~Zoupanos, Phys.~Lett.~{\bf
  B427} (1998) 291; T.~Kobayashi et al., in proc. of 
"Supersymmetry, Supergravity and Superstrings", p. 242-268, Seoul 1999.

\bibitem{pole}N.~Gray, D.~J.~Bradhurst, W.~Grafe and K.~Schilcher,
Z. Phys. {\bf C48} (1990) 673. 
 
\bibitem{ds2} Y.~Daikoku and D.~Suematsu, Prog. Theor. Phys. {\bf 104}
(2000) 827 (hep-ph/0003206).

 \end{thebibliography}
 \end{document}